# Parametric study on hydrothermal properties of condensing flow in microtube heat exchangers using numerical and experimental approaches

Abdolali K Sadaghiani [1,2,*]

[1] Faculty of Engineering and Natural Sciences (FENS), Sabanci University, Istanbul, Turkey

[2] Sabanci University Nanotechnology and Application Center (SUNUM), Sabanci University, Istanbul, Turkey

* abdolali@sabanciuniv.edu

**Abstract**

Microchannels have increasingly been used to miniaturize heat transfer equipment, improve energy efficiency, and minimize heat transfer fluid inventory. A fundamental understanding of condensation in microscale will yield far-reaching benefits for the different areas of industry. In this study, microtubes with inner diameters of 250, 500, 600, and 900 µm were used to investigate the effect of microtube diameter, inlet quality, and mass flux on the liquid/vapor interface near the wall boundaries in condensing flow. After validation with the experimental results, a transient numerical model (based on the volume of fluid approach) was developed to investigate the hydrothermal properties of condensing such as bubble dynamics, flow map transitions, transient interface shear force, and temperature on flow condensation performance in terms of heat transfer coefficient and pressure drop. The liquid film thickness, slug velocity, and location of transition from annular flow to slug flow inside the microtube were characterized for different microtubes, and the resultant alteration in condensation flow heat transfer and pressure drop is discussed in detail. The obtained results indicated that the interfacial characteristics of condensing flow in microtubes with hydraulic diameters lower than 500µm are majorly different from those with D>500µm.

**Keywords**: Flow condensation; Microtube heat exchange; Parametric study; Liquid/vapor interface; Liquid film thickness, Bubble dynamics; Flow pattern.

## 1. Introduction

Due to the high compactness needed for modern heat transfer systems, microtubes and microchannels are widely used in different areas of the industry such as heat pump, chemical engineering industry, condensers and evaporators, heating ventilating, and air conditioning systems. With advancement in nanotechnology, surface modification methods have been proposed to increase the phase change efficiency in mini and micro domains [1-5]. Effect of different parameters such as surface wettability [6,7], surface structure [8,9], surface inclination [10,11], saturation temperature [12], and tube geometry [13,14] on condensation heat transfer in conventional scale have been investigated extensively. Although many studies have been performed on two-phase heat transfer involving boiling and evaporation in microchannels, there are fewer studies on condensation phase change systems in micro domain. Coleman and Garimella presented the first studies on condensation in microchannels in 1998 [15-17]. Wongwises and Polsongkram [18] studied condensation heat transfer and pressure drop of HFC-134a in tube-in-tube heat exchangers. They investigated the effect of mass flux and condensation temperature on heat transfer and pressure drop, and new correlations were developed for heat transfer coefficient and pressure drop. Garimella et al. [19] studied condensation heat transfer in rectangular microscale geometries. Heat transfer coefficient and pressure drop of refrigerant R134a condensation were obtained in rectangular microchannels with hydraulic diameters ranging $100 < D_h < 160$ μm. They showed that as $T_{sat}$ decreases, due to the decrease in the vapor to liquid density ratio, void fraction increased which led to an increase in condensation heat transfer coefficient. Quan et al. [20] investigated steam annular flow heat transfer in trapezoidal silicon microchannels with hydraulic diameters of 127 and 173μm and proposed a semi-analytical model for liquid thin film thickness based channel dimensions and vapor quality. They concluded that the annular heat transfer coefficient in trapezoidal microchannels increases mass flux and quality but decreasing with the hydraulic diameter. Chen et al. [21] performed a visualization study on steam condensation in triangular microchannels with hydraulic diameters of 100 and 250μm, and found out that the droplet, annular, injection and slug-bubbly are the four dominant flow patterns in triangular silicon microchannels. Their results indicated that the location of injection flow moves toward the microchannel with increasing the mass flow rate. Wu et al. [22] investigated steam condensation in rectangular microchannels with hydraulic diameters of 77.5, 93, and 128.5μm, and concluded that the condensation heat transfer coefficient and pressure drop increases as the injection flow moves toward the microchannel outlet.

Fieg and Roetzel[23] conducted experimental work to calculate the laminar film condensation in/on the inclined elliptical tube. They showed that the elliptical geometry raised the heat transfer coefficient. Hussein et al.[24] investigated laminar film condensation heat transfer inside an inclined heat pipe in a solar collector. They reported that pipe inclination angle had a significant effect on condensation heat transfer inside inclined heat pipes. Fiedler and Auracher[25] presented an experimental and theoretical study on reflux condensation inside a 7mm diameter inclined tube. They developed correlations for film thickness and mean heat transfer coefficient. Akhavan-Behabadi et al.[26] investigated the combined effect of the micro-fin tube and tube inclination. They showed that the effect of the inclination angle was more eminent at low mass fluxes and vapor quality. A correlation was also developed to predict the condensation heat transfer coefficient at different vapor qualities and mass velocities.

Numerical analysis is a powerful tool for investigating the interface characteristics in microscale domains[27-33]. Ganapathy et al.[34] proposed a numerical model and investigated the R134a condensation heat transfer and flow characteristics in a rectangular microchannel with hydraulic diameter of 100µm. Zhang and his coworkers[35] investigated the condensation heat transfer and pressure drop in micro and minitubes with hydraulic diameters of 0.25, 1, and 2mm using Volume of Fluid (VOF), and concluded the existence of shear dominated flow regime at tubes of smaller diameter. Zhao et al.[36] investigated the R32 flow condensation properties of printed circuit heat exchanger having semicircular cross section with hydraulic diameter of 910µm. They discovered that smooth-annular, wavy-annular, slug and bubbly flow patterns are formed flow patterns in the microchannel. Chen et al.[37] investigated the FC-72 flow condensation in a rectangular microchannel with a 1-mm hydraulic diameter using the VOF model. The authors concluded that the formation of necks due to waves along the interface, locally increase the vapor velocity. It was found that due to decrease in interfacial area, the condensation rate deteriorated and results in reduction in the velocity of separated large bubbles from the vapor column.

The effect of non-condensable gases on steam condensing flow in a square channel with hydraulic diameter of 3mm was investigated by Vyskocil et al.[38]. The authors show that the presence of non-condensable gases deteriorate the heat transfer by forming a layer through which the steam should diffuse. Lie et al.[39] performed numerical and experimental studies on FC-72 flow condensation in a square channel with hydraulic diameter of 1mm. The authors confirmed the presence of smooth-annular, wavy-annular, transition, slug, bubbly, and pure liquid flow patterns in the minichannel. El Mghari et al.[40] performed numerical analysis on different non-circular microchannels with hydraulic diameters of 110 and 250µm. The authors concluded that compared to rectangular and triangular microchannels, square microchannel have lower heat transfer coefficients. Furthermore, it was shown that the average condensation heat transfer coefficient increases with aspect ratio for rectangular microchannels. Szijártóa et al.[41] investigated the steam condensation by implementing three different condensation models into VOF method. The authors concluded that phase field model provided more accurate results compared to other models.

Majority of studies on condensing flows in microchannels with a hydraulic diameter smaller than 1mm have used silicon microchannels, which is not applicable for many practical engineering applications. On the other hand, the literature lacks a parametric study on the effect of channel dimension on hydro-thermal properties of condensation in microchannel. To fill these gaps, this study investigates the condensation heat transfer and pressure drop in microtubes with inner diameters of 250, 500, 600, and 900 µm using numerical and experimental approaches. Steam was used as the working fluid with inlet mass velocities in the range of 25 to 225 kg/m$^2$.s. Using a data acquisition system, the variation of wall temperatures and pressure drops were obtained experimentally. The local and average heat transfer coefficient and pressure drop in microtubes were evaluated and the obtained results were discussed. Using numerical analysis, the effect of microtube diameter on flow regime and bubble dynamics were investigated.

## 2. Numerical modeling
### 2.1 Volume of Fluid Model (VOF Model)

There are several models proposed for multiphase modeling including Volume of Fluid (VOF)[42] and Eulerian Models [28, 43]. The VOF model can model two or more immiscible fluids by solving a single set of momentum equations and tracking the volume fraction of each of the fluids throughout the domain [44]. The VOF model differentiates two separate fluids by using a scalar indicator function between zero and one. Since one value is required to be accorded to each mesh cell and one scalar convective equation is solved to propagate the indicator function through the computational domain, generally this approach is more economical (computational cost) relative to other methods. The VOF formulations in ANSYS rely on the non-interpenetrating behavior of phases, where the summation of volume fractions in each cell is unity. The field variables are shared by all phases and is represented by a volume average value, meaning that the representative properties in any given cell correspond for one phase or mixture of phases depending on their volume fraction values. In other words, if the volume fraction of q$^{th}$ phase is shown by $\alpha_q$, in any cell $\alpha_q = 0$ means the cell is empty, $\alpha_q = 1$ means the cell is full of the q$^{th}$ phase, and $0 < \alpha_q < 1$ means the cell contains the interface between the q$^{th}$ phase and the other phases) [44].

### 2.2 Numerical Formulation

Using the VOF model, the conservations equations were solved for the multiphase flow inside the computational domain.

### 2.2.1 Conservation of mass and phase change model:

The conservation of mass equation for the vapor phase is given by eq. 1, where the mass transfer mechanism between the vapor phase and the liquid phase is given on the right-hand side of the equation. The widely used Lee model [45] was used for condensation mass transfer (source term of $\dot{m}_{vl}$) from vapor phase (subscript v) to liquid phase (subscript l) is given by eq. 2 [34, 46].

$$\alpha_v + \alpha_l = 1 \tag{1}$$

$$\frac{\partial(\alpha_v \rho_v)}{\partial t} + \nabla \cdot (\alpha_v \rho_v \vec{V}_v) = S_v \tag{2}$$

$$\frac{\partial(\alpha_l \rho_l)}{\partial t} + \nabla \cdot (\alpha_l \rho_l \vec{V}_l) = S_l \tag{3}$$

$$S_v = -S_l = \frac{\beta_{pf}(T_{cell} - T_s)|\nabla \alpha|}{h_{lv}}$$

$$\beta_{pf} = \frac{6\sqrt{2}}{5} \frac{D_l \rho_l C_{pl}}{w} \tag{4}$$

The β$_{pf}$ coefficient is calculated using the liquid characteristics ($D_l = \lambda_l / (C_{p,l} \rho_l)$) and interface characteristics length ($\omega = 2\delta y$). Basically, the driving mechanism for liquid/vapor phase change is the temperature gradient between the saturation temperature and interfacial cells. As can be seen, the interface length is approximated by the cell height and interfacial cells are cells that experience abrupt change in the void fraction. It should be noted that the saturation temperature is given as a function of local pressure. The value of volume fractions in each cell is between 0 and 1, and the summation of volume fraction in each cell is equal to unity as shown in Eq. 1.

### 2.2.2 Conservation of Momentum

In the VOF model, a single momentum equation is solved in the computational domain, resulting in a shared velocity field among the phases. Therefore, the momentum equation is dependent on the phase's volume fraction. The momentum equation is as follows:

$$\frac{\partial(\rho\vec{V})}{\partial t} + \nabla \cdot (\rho\vec{V}\vec{V}) = -\nabla P + \nabla \cdot \left[\mu\left(\nabla\vec{V} + \nabla\vec{V}^T\right)\right] + \rho\vec{g} + \vec{F}_s \quad (5)$$

Here, $\rho = \alpha_v \rho_v + \alpha_l \rho_l$ is the volume-average density, $\mu = \alpha_v \mu_v + \alpha_l \mu_l$ is the volume-average viscosity ($\mu$), V is the shared velocity profile, P is static pressure, and $\vec{F}_s$ is the surface tension force. Although the surface tension is a surface force, in the commercial software it is converted to the volume force using the continuum surface force (For more information refer to Brackbill et al. [47]). The surface tension force is calculated as follows:

$$\vec{F} = 2\sigma \frac{\alpha_l \rho_l \kappa_v \nabla \alpha_v + \alpha_v \rho_v \kappa_l \nabla \alpha_l}{\rho_l + \rho_v} \quad (6)$$

$$\kappa_v = \nabla \cdot \frac{\nabla \alpha_v}{|\alpha_v|} \quad (7)$$

Here, $\kappa_v$ is the vapor phase curvature which is represented by the divergence of the unit normal vector ($\kappa_v = \nabla \cdot \frac{\nabla \alpha_v}{|\alpha_v|}$). It should be noted that when both phases are presented in the cell, $\kappa_v = -\kappa_l$ and $\nabla \alpha_v = -\nabla \alpha_l$.

### 2.2.3 Conservation of Energy:

A single energy conservation equation is solved for two-phase flow with modified thermophysical properties. The conservation of energy is given as follows:

$$\frac{\partial(\rho E)}{\partial t} + \nabla \cdot \left[\vec{V}(\rho E + p)\right] = \nabla \cdot (k_{eff} \nabla T) + S_E \quad (8)$$

Here, $E = \frac{(\alpha_v \rho_v E_v + \alpha_l \rho_l E_l)}{\alpha_v \rho_v + \alpha_l \rho_l}$, T, and $K_{eff}$ are the energy, the temperature, and the effective thermal conductivity, respectively. The latent heat associated with the phase change phenomena was modeled as an energy source and was represented by the term $S_E$ as follows:

$$S_E = S_l \left(h_{fg} - E_{correction}\right) \quad (9)$$

The correction terms ($E_{correction}$) is considered in the energy equation to compensate the large property contrast between the liquid and vapor phases as follows [48] [41]:

$$E_{correction} = \left(\rho_l C_{p,l} - \rho_v C_{p,v}\right) \frac{w}{\sqrt{2}} \left(\frac{1}{\rho_v} - \frac{1}{\rho_l}\right) \vec{n}_l \cdot \vec{\nabla T} \quad (10)$$

Table 1 shows some of the available correlations used as the source terms for the energy and mass conservation equations.

Table 1. Some of the available source terms

| Phase change model | Mass and energy source terms | |
|---|---|---|
| Zhang et al. [49] <br> In this model, the interface temperature is assumed by adapting a huge heat source term. | $S_v = -S_l = -\rho_v \left( \dfrac{\partial \alpha_v}{\partial t} + \nabla.(\alpha_v \vec{v}) \right)$ <br><br> $S_E = 10^{10} c_p (T_{sat} + T - 2T_{ref})$ | (11) |
| Yuan et al. [50] <br> This model assumed the interfacial temperature by adapting a huge heat source term. | $S_v = -S_l = -\rho_v \left( \dfrac{\partial \alpha_v}{\partial t} + \nabla.(\alpha_v \vec{v}) \right)$ <br><br> $S_E = 10^{10} (T_{sat} - T)$ | (12) |
| Aghanajafi et al. [51] <br> In this model, the energy balance is assumed by Fourier's law in the liquid/vapor interface. | $S_v = -S_l = -\dfrac{(\alpha_v \lambda_v + \alpha_l \lambda_l)(\nabla \alpha_l . \nabla T)}{h_{fg}}$ <br><br> $S_E = -(\alpha_v \lambda_v + \alpha_l \lambda_l)(\nabla \alpha_l . \nabla T)$ | (13) |
| Sun et al. [46] <br> This model assumes energy balance and is suitable for both saturated and unsaturated conditions. | $S_v = -S_l = \dfrac{2\lambda_l (\nabla \alpha_l . \nabla T)}{h_{fg}}$ <br><br> $S_E = -2\lambda_l (\nabla \alpha_l . \nabla T)$ | (14) |
| Rattner and Garimella [52], and Onishi et al [53] <br> This model assumes that the internal heat transfer resistance is neglected. | $S_v = -S_l = -\dfrac{\rho c_p (T_{sat} - T)}{h_{fg}.\Delta t}$ <br><br> $S_E = \dfrac{\rho c_p (T_{sat} - T)}{\Delta t}$ | (15) |
| Lee [54], De Schepper et al. [55], Liu et al. [56] <br> This model is simplified from the Hertz–Knudsen equation. | $S_v = -S_l = -f_e \alpha_v \rho_v \dfrac{(T_{sat} - T)}{T_{sat}}$ <br><br> $S_E = f_e \alpha_v \rho_v h_{fg} \dfrac{(T_{sat} - T)}{T_{sat}}$ | (16) |

It should be noted that the mass and energy source terms were not part of the CFD code and were programmed into the numerical model by defining additional routines. The gradients of temperature and volume fraction were calculated using the divergence theory and were used to determine the three source terms given by conservation equations.

### 2.2.4 Turbulence model:

A range of turbulence models has been implemented in flow condensation simulations. Shear Stress Transport (SST) model has been used by many researchers for computational condensation phase change analysis [39, 57, 58]. The SST k-ω is a two-equation eddy-viscosity model, which assumes the proportional relationship between Reynolds shear stress and turbulent kinetic energy in the boundary layer [59]. In this model, k and ω are the turbulence kinetic energy and specific dissipation rate, respectively, where the turbulent viscosity is limited to tackle the transport of turbulent shear stress. The low-Reynolds number form the SST k-ω model are given as follows:

$$\frac{\partial(\rho k)}{\partial t}+\frac{\partial}{\partial x_i}(\rho k v_i) = \frac{\partial}{\partial x_j}\left(\left(\mu+\frac{\mu_t}{Pr_{t,k}}\right)\frac{\partial k}{\partial x_j}\right)+\mu_t S^2 - \rho\beta^* k\omega \quad (17)$$

$$\frac{\partial(\rho \omega)}{\partial t}+\frac{\partial}{\partial x_j}(\rho \omega v_j) = \frac{\partial}{\partial x_j}\left(\left(\mu+\frac{\mu_t}{Pr_{t,\omega}}\right)\frac{\partial \omega}{\partial x_j}\right)+\frac{a}{v_t}\mu_t S^2 - \rho\beta^*\omega^2$$
$$+2(1-F_1)\rho\frac{1}{1.168\omega}\frac{\partial k}{\partial x_j}\frac{\partial \omega}{\partial x_j}+S_\omega \quad (18)$$

Here, $Pr_{t,k}$, $Pr_{t,\omega}$, $\mu_t$, $S$, $v_t$ and $a$ are turbulent Prandtl numbers for k and ω, respectively, the turbulent viscosity, the strain rate, the kinematic viscosity, and the coefficient associated with the production of ω, respectively. $\beta^*$ and $S_\omega$ are calculated as follows:

$$\beta^* = 0.09\left(\frac{4/15+(Re_t/8)^4}{1+(Re_t/8)^4}\right)(1+1.5F(M_t)) \quad (19)$$

$$S_\omega = 2\alpha_i|\nabla\alpha_i|\Delta n\beta\rho_i\left(\frac{f6\mu_i}{\beta\rho_i\Delta n^2}\right)^2 \quad (20)$$

$F(M_t)$ is the compressibility function, and the turbulent Reynolds number defined as:

$$Re_t = \frac{\rho k}{\mu\omega} \quad (21)$$

## 2.3 Computational Domain and solution procedure:

### 2.3.1 Computational domain:

The computational domain consisting of a microtube with a length of 9 cm and the inner diameter of 500 µm is displayed in Figure 1. A velocity boundary condition was specified at the channel inlet and a pressure-based boundary condition was specified at the channel outlet. No-slip and constant heat flux were specified as the wall hydraulic and thermal boundary conditions, respectively. To simplify the analysis, the microtube walls were not modeled, and therefore the effects of transient conduction within the wall not considered in the present study [27].

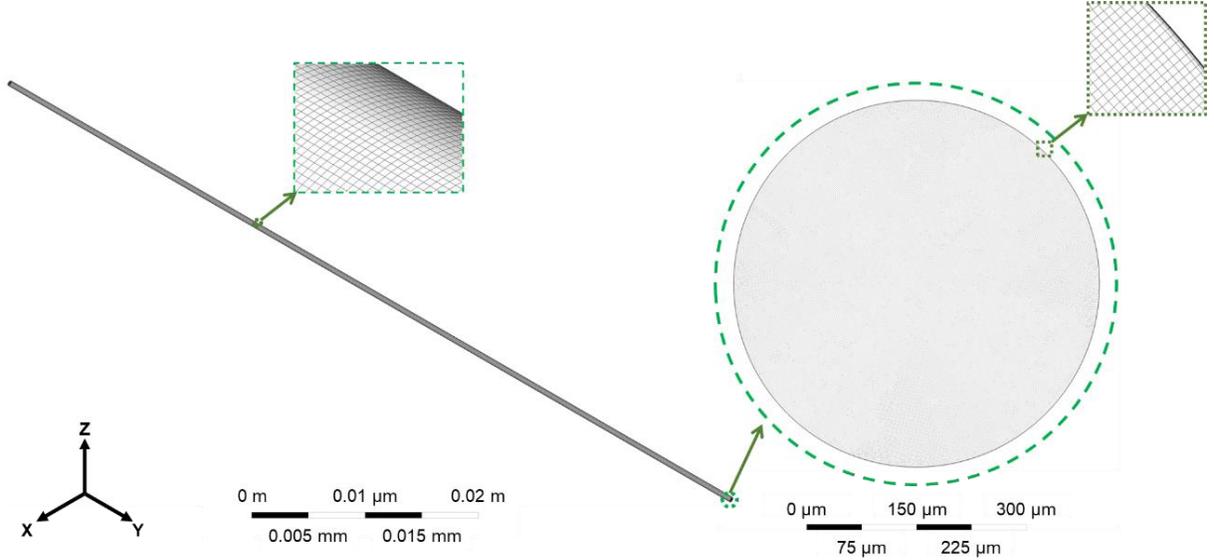

Figure 1. The computational domain for microtube with a hydraulic diameter of 500 μm and length of 9 cm.

### 2.3.2 Grid dependency study

The mesh dependency study for each case was performed by investigating the change in wall temperature at four locations of 1, 2, 3, 4, and 5mm for four different grid sizes in the range of 2μm to 10μm. Figure 2 shows the variation in the calculated results for the grid sizes of 2, 5, 8, and 10μm. Here, a microtube with a diameter of 500μm, a length of 9cm, and boundary conditions of the inlet mass flux of 100 kg/m$^2$.s and the wall heat flux of 30 kW/m$^2$ was used for grid dependency investigation. Since the difference in wall temperatures between cases with the mesh size of 2μm and 5μm is less than 1%, for the rest of the modeling, the minimum grid size of 5μm is used to reduce computational cost. To capture the thin film during the condensation (especially for annular and slug flow regimes), refined meshed were used near the channel walls. A sensitivity analysis was performed to show the effect of wall inflation on wall shear stress. Accordingly, it was found that the difference in the wall shear stress between the cases with inflation layers of 15 and 20 is less than 1%. Therefore, 15 inflation layers with a growth rate of 1.05 is used for the rest of the numerical analysis. To ensure that the first near-wall grid is sufficiently small, mesh refinement near the wall boundary was evaluated using y$^+$ non-dimensional, which is defined as follows:

$$Y^+ = \frac{yV_f}{v} \quad (22)$$

Here, y is the normal distance to the wall, $V_f = \sqrt{\tau_w/\rho}$ is the friction velocity, $\tau_w$ is the wall shear stress, ρ is the density, and $v$ is the kinematic viscosity. In the current numerical analysis, the $y^+$ values were lower than 2.29 for all cases, which meet the conditions of accurate boundary layer sensitivity.

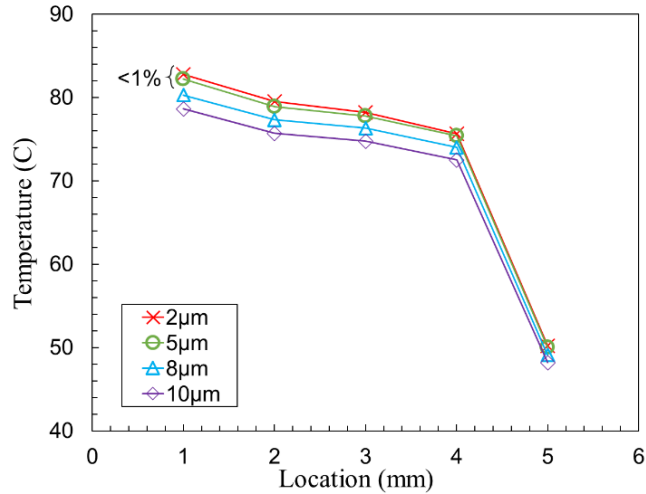

Figure 2. Grid dependency test for the mass flux of 100 kg/m$^2$.s

### 2.3.3 Solution procedure

ANSYS Fluent (2019R2 – Ansys, Inc.) commercial software was used to implement the numerical formulations. The Finite Volume Method is used to discretize the governing equations. The Explicit Volume of Fluid (VOF) and Geometric Reconstruct schemes were used to capture the liquid/vapor interface. The source terms mentioned in the conservation equations were calculated using UDF and hooked to their corresponding transport equations. An implicit Body Force formulation was utilized for the partial equilibrium in pressure gradient and surface tension force. The gradient of scalar in discontinuities was calculated using Green-Gauss Cell-based scheme. The second-order Upwind and Implicit schemes were used for spatial and temporal equations, respectively. The Pressure-Implicit with Splitting of Operators (PISO) method was used for the pressure-velocity coupling.

### 3. Experimental Setup and Data Reduction

### 3.1 Experimental setup and test section

Schematics of the experimental setup and test section are shown in Figure 3. The experimental setup consists of the main loop (condensation) and cooling loop. The main loop includes the filters, gear pump, valves, flow meters, temperature, and pressure sensors. The cooling loop was used to provide heat removal from the test section. The test section is a shell and microtube heat exchanger. The inner tube is a microtube with a different inner diameter, where the outer tube has an inner diameter of 3mm. Figure 4 shows the water contact angle measurements and SEM images of the utilized microtubes. The heat removal rate was regulated by changing the cooling flow rate.

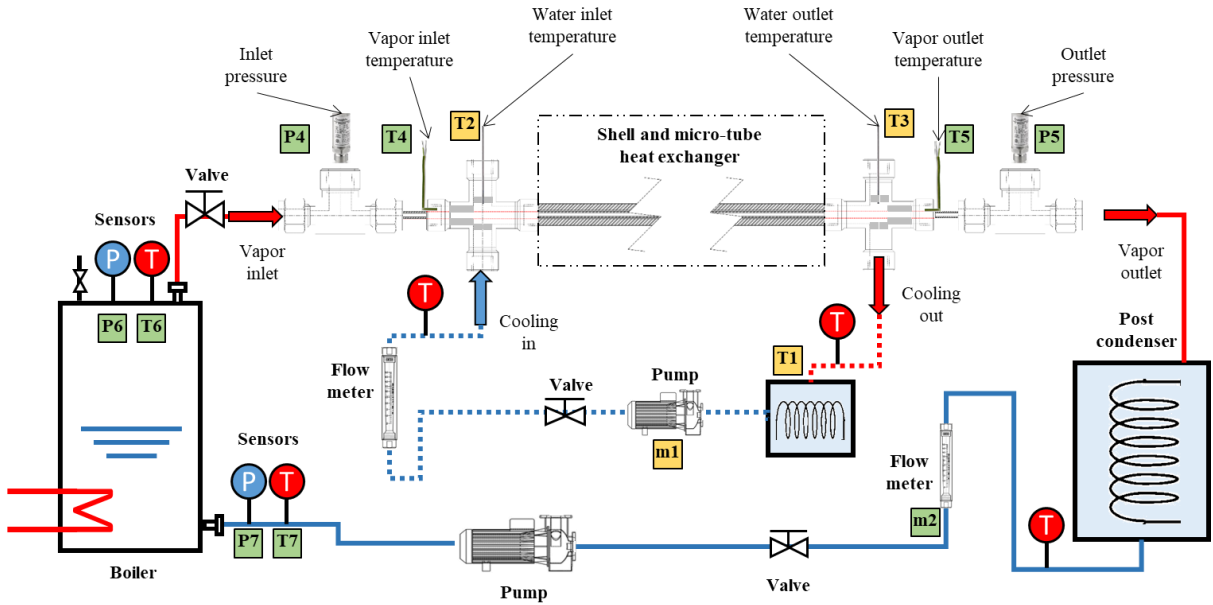

Figure 3. Schematic of the experimental setup and shell and microtube heat exchange

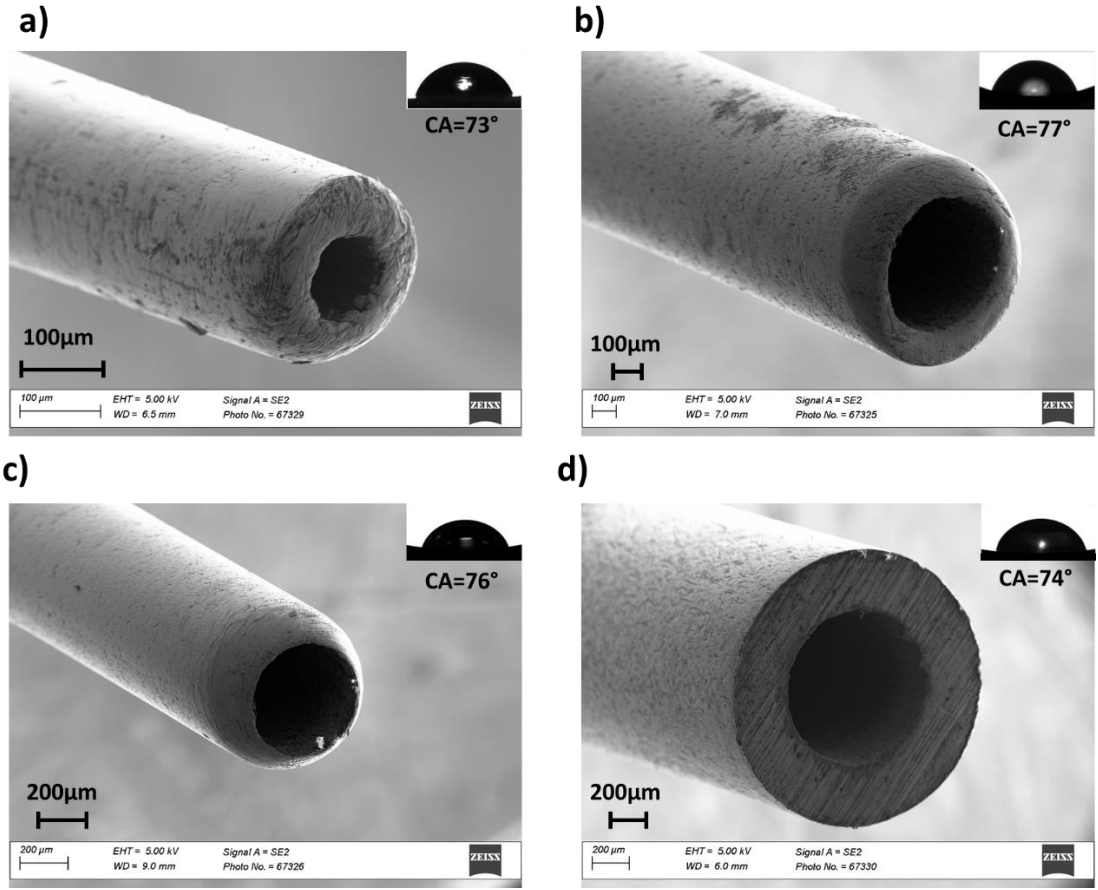

Figure 4. Water contact angle measurements and SEM images of microtubes with inner diameters of a) 250μm; b) 500μm; c) 600μm; d) 900μm.

## 3.2 Data Reduction

Before performing the condensation experiments, the efficiencies of the evaporator and condenser were obtained by calibrating their thermal performance. The calibration experiment was performed under single-phase convective heat transfer conditions [34, 60]. The thermal efficiency of the evaporator is calculated using the following equation.

$$\eta_{eva} = \frac{m_w c_{p,w} \left( T_{w,in} - T_{w,out} \right)}{UI} \tag{23}$$

where $m_w$, $c_{p,w}$, $T_{w,in}$, $T_{w,out}$, $U$ and $I$ are the mass flow rate and specific heat of water, the water temperatures at the evaporator outlet and, water temperatures at the evaporator inlet (correspond to the $T_7$ and $T_4$ shown in Figure 3, respectively), the voltage and current, respectively. According to the obtained results, the average thermal efficiency of the evaporator was obtained as 0.89.

A similar procedure was performed to obtain the condenser efficiency as follows:

$$\eta_{con} = \frac{m_{w,s} c_{p(w,s)} \left(T_{(w,s),in} - T_{(w,s),out}\right)}{m_{w,t} c_{p(w,t)} \left(T_{(w,t),in} - T_{(w,t),out}\right)} \quad (24)$$

Here the subscripts (w,t) and (w,s) stand for the water on the tube side and the water on the shell side of the shell-tube heat exchanger (condenser). The measured average condenser efficiency is about 0.98.

The condenser inlet enthalpy ($i_{r,in}$) and quality ($x_{in}$), stem from the evaporator efficiency:

$$i_{r,in} = i_{r,1} + \frac{UI \eta_{con}}{m_r} \quad (25)$$

$$x_{in} = \frac{i_{r,in} - i_{l,in}}{i_{lg,in}} \quad (26)$$

where $i_{r,1}$ is the water's enthalpy at the evaporator inlet, $i_{l,in}$ and $i_{lg,in}$ are the saturated liquid enthalpy and latent heat of evaporation based on the inlet pressure, respectively.

The condenser outlet parameters were obtained in terms of the condenser efficiency. Heat received by the cooling water in the shell tube is given as:

$$Q = m_m C_{p,c} \left(T_{c,out} - T_{c,in}\right) \quad (27)$$

Heat flux based on the inner copper tube wall surface is expressed as:

$$q = \frac{Q}{\pi d_i L_e} \quad (28)$$

where $L_e$ is the effective heat transfer length. The outlet mixture enthalpy and quality are calculated as follows

$$i_{r,out} = i_{r,in} - \frac{Q}{m_r \eta_{con}} \quad (29)$$

$$x_{out} = \frac{i_{r,out} - i_{l,out}}{i_{lg,out}} \quad (30)$$

Here, $i_{l,out}$ and $i_{lg,out}$ are the saturated liquid enthalpy and latent heat of evaporation based on the outlet pressure, respectively. It should be noted that the inlet and outlet vapor qualities are calculated based on the corresponding pressures. The condensation heat transfer coefficient h is based on the average vapor mass quality, stated as $x_{ave} = (x_{in} + x_{out})/2$. Thus, $h_{tp}$ is calculated as [61]:

$$h_{tp} = \frac{1}{\frac{1}{h_{to}} - \frac{d_i}{2k_w} \ln\left(\frac{d_{outer}}{d_{inner}}\right) - \frac{d_{inner}}{d_{outer}} \frac{1}{h_c}} \tag{31}$$

where $k_w$ is the thermal conductivity of the microtube, $h_{to}$ is the total heat transfer coefficient, and $h_c$ is the heat transfer coefficient of cooling water. The calculation of $h_{to}$ and $h_c$ can be found in Ref [62]. Pressure drops were correlated based on the Chisholm method, where the two-phase multiplier ratio is expressed as:

$$\varphi_l^2 = \frac{\Delta P_{Tf,f}}{\Delta P_{sp,f}} \tag{32}$$

where $\Delta P_{Tf,f}$ and $\Delta P_{sf,f}$ are the two-phase friction pressure drop and friction pressure drop if liquid flows alone. The Lockhart-Martinelli parameter X is calculated as:

$$X^2 = \frac{\Delta P_{liquid,f}}{\Delta P_{gas,f}} \tag{33}$$

where $D_{gas,f}$ is the friction pressure drop if the vapor phase flows alone. Finally, the following expression is included to correlate the two-phase friction pressure drop.

$$\varphi_l^2 = 1 + \frac{C}{X} + \frac{1}{X^2} \tag{34}$$

## 4. Results and discussion

### 4.1 Hydrodynamic analysis

Figure 5 shows the obtained flow patterns for the microtube with a hydraulic diameter of 250μm. Four major flow patterns were observed. As the flow passes through the microtube, the flow pattern changes from annular flow to injection flow, slug flow, and to bubbly flow along the microtube. As can be seen, the liquid film thickness in the annular flow regime increases in the flow direction. At sufficiently downstream of the annular flow, the liquid films at the wall boundaries start to bridge (yellow arrows in Figure 5a), which results in a severe increase in vapor flow velocity. The velocity difference between liquid film and vapor flow results in an interfacial instability in this region. At some point, the surface tension force is not sufficient enough to overcome this instability, and a break-up takes place inside the microtube (Figure 5b). The frequency of the bubble break-up depends on the flow velocity and wall thermal conditions. The break-up frequency increases with wall heat flux and decreases with the vapor mass flux. The separated vapor produces Taylor bubbles form the slug flow regime inside the microtube (Figure 5a). As slugs move toward the outlet, heat transfer through the microtube walls gradually change the slugs to spherical bubbles and results in the formation of bubbly flow. Eventually, the spherical bubbles condense into the liquid flow and form single-phase liquid flow. It should be noted that decreasing the inlet mass flux, increasing the inlet vapor quality, and decreasing the wall heat flux have similar effects on the transition between flow patterns.

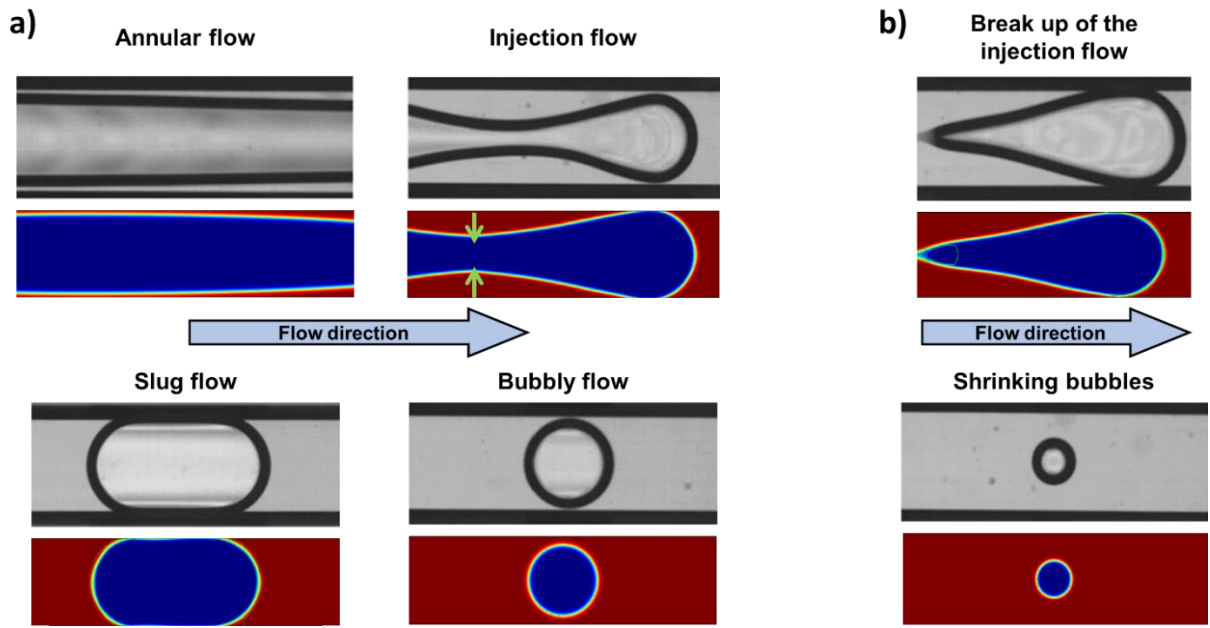

Figure 5. The numerical analysis and corresponding experimental results [63] of microtube with D=250µm a) predicted flow patterns; b) break-up of the injection flow and shrinking of the bubbles in the stream

Condensation is driven by the temperature gradient at the liquid/vapor interface. The local condensation heat transfer in a laminar condensing flow is solely dependent on the liquid film thickness (δ). Therefore, development of interface along the microtube is crucial for thermal performance of a condensing flow, especially in microtube. At very low vapor qualities, the vapor velocity and interfacial stress are almost zero. At this condition, the liquid film thickness is almost equal to microtube diameter (D) and single-phase liquid flow exists in the microtube. On the other hand, at vapor qualities close to one, the liquid velocity is very low and interfacial stress is almost equal to the wall shear stress. At this condition, the liquid film thickness becomes zero, and single-phase gas flow exists in the microtube [64].

Figure 6a shows the average liquid film thickness and dimensionless liquid film thickness (δ/D) versus vapor quality for different microtubes and mass fluxes. As can be seen, the liquid film thickness decreases with vapor quality in both microtubes. It was found that the effect of vapor quality on liquid film thickness is more pronounced at smaller diameters, which indicates the dominancy of interfacial surface tension force at smaller scales. Furthermore, larger values if non-dimensional liquid film thickness (δ/D) in larger microtubes decreases the effective heat transfer from wall to the two-phase interface, which in turn is the major reason for condensation heat transfer deterioration at microtubes with larger diameters. The location of transition from annular flow to slug flow is shown in Figure 6b. A linear relationship exists between flow transition position and inlet vapor mass flux. For the same inlet mass flux, microtube diameter decrement results in the extension of transition location towards the outlet. Similar observations are reported in the literature [21, 65].

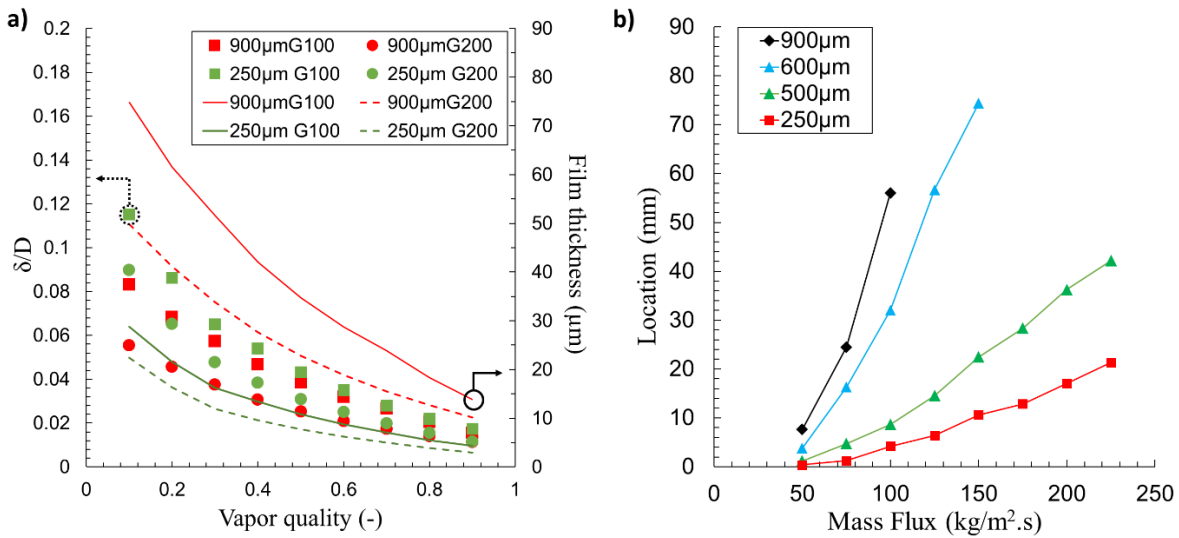

Figure 6. a) Liquid film thickness and non-dimensional liquid film thickness at different vapor qualities for steam mass flux of 150 kg/m$^2$.s and cooling heat flux of 30 kW/m$^2$; b) Location of annular flow to slug flow pattern transition

Increasing the inlet mass flux or decreasing the wall heat flux result in the expansion of the vapor column. This is indicated in Figure 7a. The vapor columns expand from the inlet to y=180mm and y=110mm for the inlet mass flux of 150 and 100 kg/m$^2$.s, respectively. Similarly, for the inlet mass flux of 100 kg/m$^2$.s, decreasing the wall heat flux from 20 to 15 kW/m$^2$ results in the vapor column elongation. Larger slugs take off from the elongated vapor columns. As a result, increasing the mass flux or decreasing the wall heat flux results in an enhancement in bubble volume. As can be seen in Figure 7a, the initial bubble volume at a wall heat flux of 20 kW/m$^2$ are 2.1 and 4.2 mm$^3$ for mass fluxes of 100 and 150 kg/m$^2$.s, respectively. Furthermore, for the mass flux of 100 kg/m$^2$.s, the initial bubble volumes are 2.1 and 1.1 mm$^3$ for wall heat fluxes of 20 and 30 kW/m$^2$, respectively. According to the obtained results, it was found that the effect of wall heat flux on bubble size increases with the inlet mass flux. Bubble velocities after taking off from the vapor slug are shown in Figure 7b. As can be seen, smaller bubbles show larger variation in the bubble size with time, especially at the first 30 milliseconds from the take-off. Upon separation from the large vapor column, the neck interface experiences a large gradient in the surface tension. This results in flow disturbances around the bubbles [37].

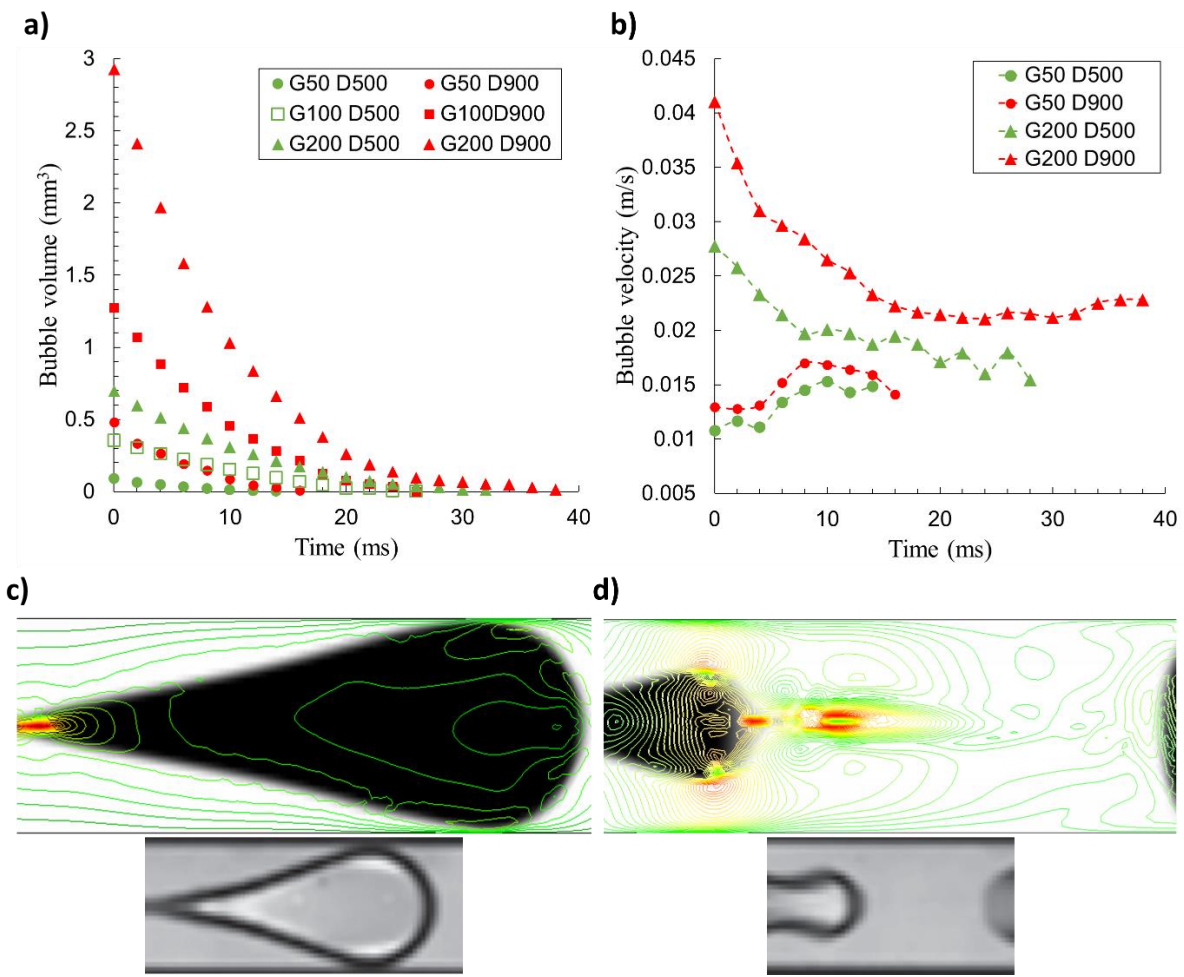

Figure 7. a) Transient bubble volume after breaking off from the vapor column; b) Transient bubble velocity after breaking from the vapor column; c) flow disturbances upon droplet break up; d) flow disturbances right after bubble break up (visual results from [66])

Figure 8 shows the local wall shear stresses in the microtube with D=500µm for G=200 kg/m².s and cooling heat flux of 30 kW/m². Here, the wall shear stress is calculated as follows:

$$\tau = -\mu \left.\frac{\partial u}{\partial r}\right|_{wall} \tag{35}$$

As can be seen, local wall shear stress at the regions near the microtube inlet are higher. Here, the presence of annular flow results in thinner liquid film thickness and larger velocity gradients. A decreasing trend in wall shear stress is evident as the liquid quality increases along the microtube. As can be seen, there are oscillations in wall shear stress in the injection flow, slug flow, and bubbly flow regimes. At injection and slug flow regimes, the bubble diameter is almost equal to the microtube, which results in thinning the liquid film at wall boundaries. As vapor size decreases along the microchannel, the liquid film thickness increases and results in wall shear stress reduction. It should be noted that the zero slopes of the wall pressure in the bubble region indicate that all pressure drops can be neglected in this region. Similar observations were reported in the literature [67-69].

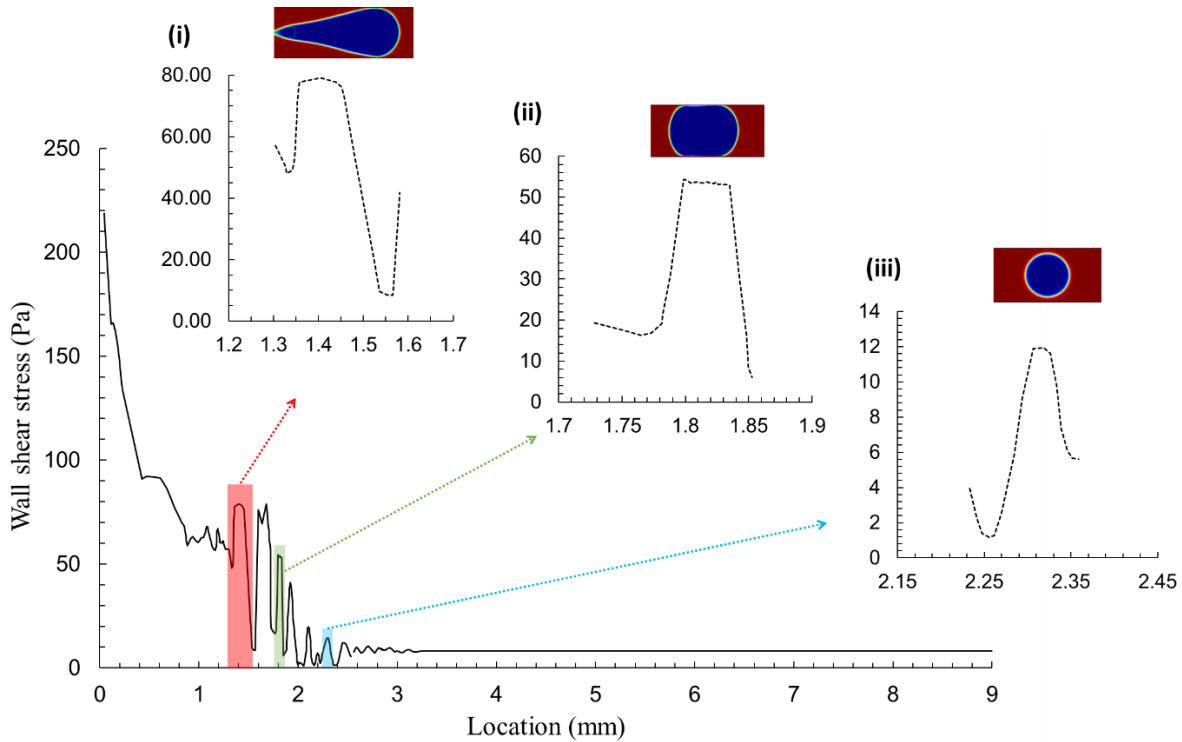

Figure 8. Wall pressure profile for microtube with D=250µm (i) bubble upon break-up, (ii) slug after break-up (slug), and (iii) single bubble at G=200 kg/m².s and cooling heat flux of 30kW/m²

In the absence of gravitational pressure drop, the total pressure drop in a multiphase flow in microtube is obtained as a summation of the frictional ($\Delta P_{Fr}$) and acceleration ($\Delta P_a$) pressure drops ($\Delta P_g$) as follows: $\Delta P_{TP} = \Delta P_a + \Delta P_{fr}$. The acceleration pressure drop due to the density change is calculated by Eq. 32. Then the frictional pressure drop is calculated by subtracting the acceleration pressure drop from the total pressure drop, as given by Eq. (34):

$$\Delta P_a = \dot{m}^2 \left\{ \left[ \left(\frac{x^2}{\alpha \rho_v}\right) + \left(\frac{(1-x)^2}{(1-\alpha)\rho_l}\right) \right]_{out} - \left[ \left(\frac{x^2}{\alpha \rho_v}\right) + \left(\frac{(1-x)^2}{(1-\alpha)\rho_l}\right) \right]_{in} \right\} \tag{36}$$

$$\Delta P_{fr} = \Delta P_{TP} - \Delta P_a \tag{37}$$

Figure 9a shows the comparison between the obtained experimental results and those predicted by equation 34. As seen for the horizontal configuration, $\phi_l^2$ based on the two-phase frictional pressure drop has a good agreement with the Chisholm method for C=12. The measured total pressure drop along the microtubes are shown in Figure 9b. The total pressure drop increases with inlet mass flux, due to larger shear stress between two-phase flow and wall. Furthermore, it is evident that the microtube diameter has an inverse effect on pressure drop.

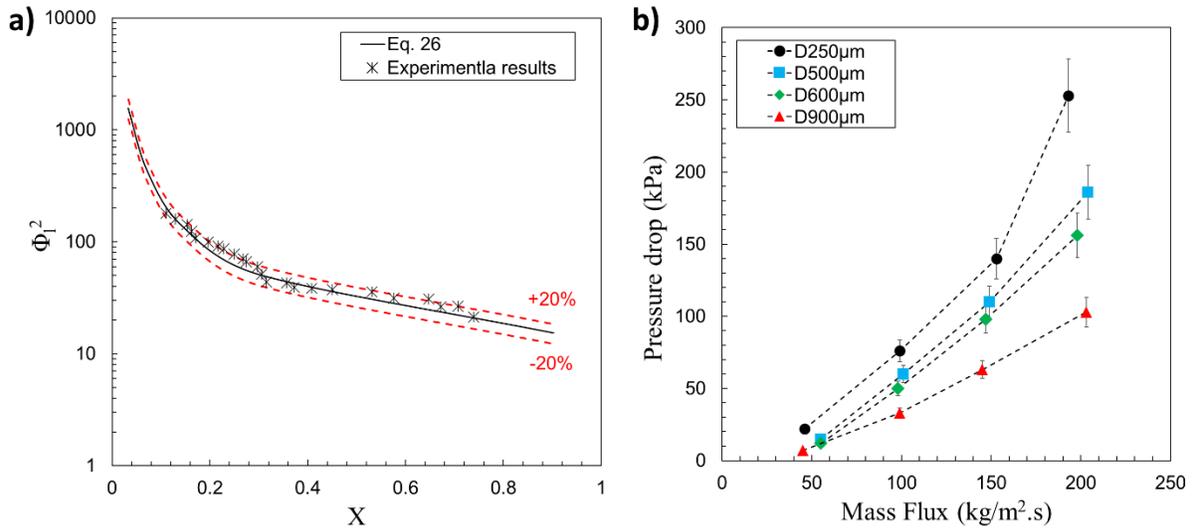

Figure 9. a) Comparison between obtained experimental results and those available in the literature; b) Calculated experimental pressure drop gradients for different case studies

## 4.2 Thermal analysis

Figure 10 shows the condensation heat transfer coefficients for microtubes versus steam mass flux in the microtubes with different hydraulic diameters while keeping the inlet water cooling temperature and flow rate at 23°C and 150 ml/min, respectively. As can be seen, the condensation heat transfer coefficient increases with inlet mass flux for all cases. Steam mass flux increment results in the extension of the vapor column in the microtube. As a result, the annular flow appears in a larger portion of the microtube. Due to the larger heat transfer coefficient of annular flow compared to slug and bubbly flow regimes, increasing the steam mass flux enhances the condensation heat transfer coefficient. However, once the annular flow occupies the entire microtube, the dependency of the heat transfer coefficient on steam mass flux decreases. The results revealed different heat transfer coefficient trends for microtubes with $D_h \leq 500\mu m$ and those with $D_h > 500\mu m$. For microtubes with $D_h \leq 500\mu m$, the effect of inlet mass flux on heat transfer coefficient decreases with steam flow rate, indicating that the convection effect on condensation heat transfer decreases with steam mass flux.

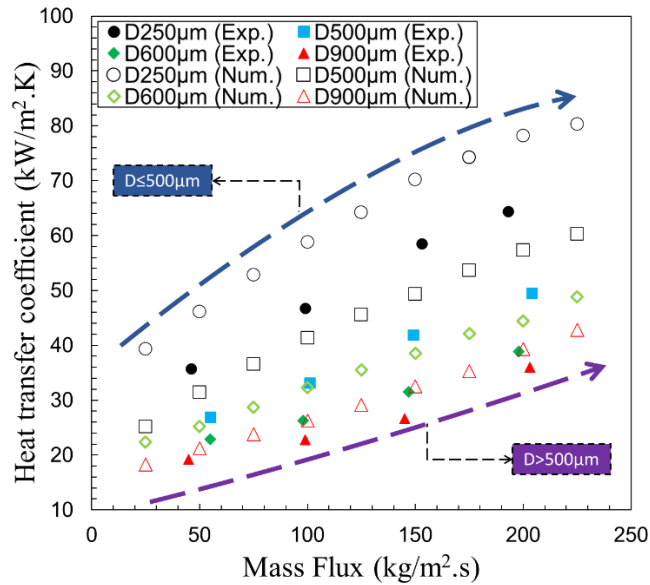

Figure 10. Numerical and experimental results of average heat transfer coefficients for different microtubes at different mass flux

The effect of vapor quality on condensation heat transfer is shown in Figure 11. As the quality increases, the vapor quality and likelihood of annular flow in the microtube increase. Since liquid film increases the thermal resistance between the vapor and the wall, the annular flow regime has a higher heat transfer rate compared to slug flow and bubbly flow regimes. As a result, the heat transfer coefficient increases with vapor quality.

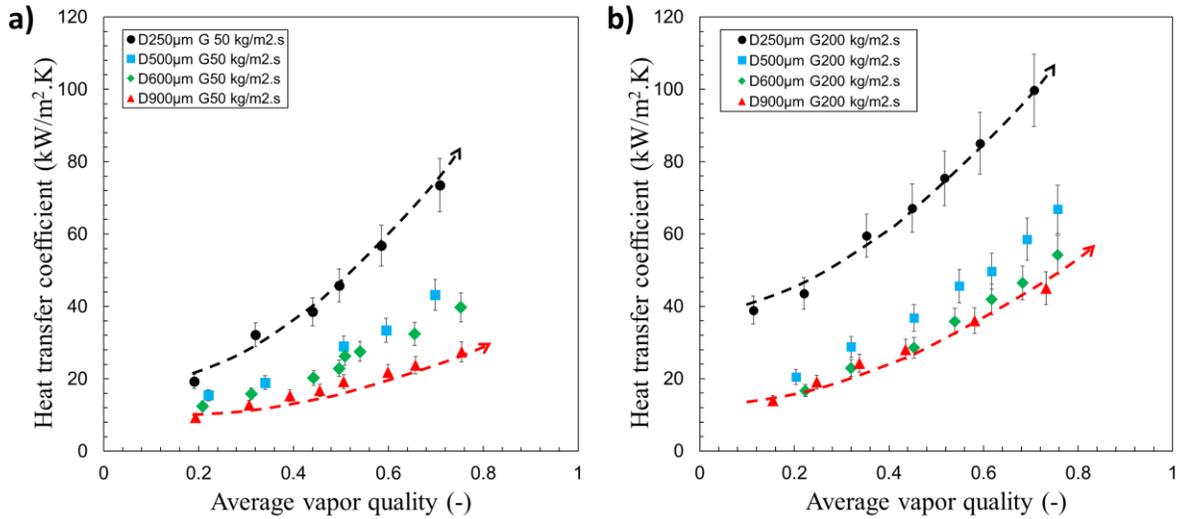

Figure 11. Obtained overall heat transfer coefficients vs. vapor quality a) Effect of hydraulic diameter at low mass flux; b) effect of hydraulic diameter at high mass flux.

The obtained experimental transient heat transfer coefficients for microtubes with the inner diameter of 500 and 600 µm within 2.5 seconds is shown in Figure 12a. As seen, the heat transfer coefficient in the microtube with an inner diameter of 500 µm is more stable compared to the microtube with a larger inner diameter. One reason might be related to the flow map transition in larger microtube. As the hydraulic diameter decreases, the annular flow regime extends inside the microtube. This results in a more stable liquid-vapor interface. The rather unstable behavior in the larger microtube might be related to the vapor-liquid interface wave and resultant temperature fluctuations. The time-averaged heat flux is defined to compare the variation in cross-sectional heat flux between different microtubes.

$$\left.\frac{\partial T}{\partial r}\right|_w = -\frac{q}{k} \quad (29)$$

$$T_{ave} = \int_{t=0}^{t=1} T|_w \, dt \Big/ \Delta t \quad (30)$$

Figure 12b shows the $T_{ave}$ along the microtubes with different hydraulic diameters. The wall temperature has a decreasing trend up to the vapor column break up point. High vapor fraction at this location could be a reason for decrees in local wall temperature. Furthermore, the wall temperate fluctuations between the inlet and break up point decreases with hydraulic diameter.

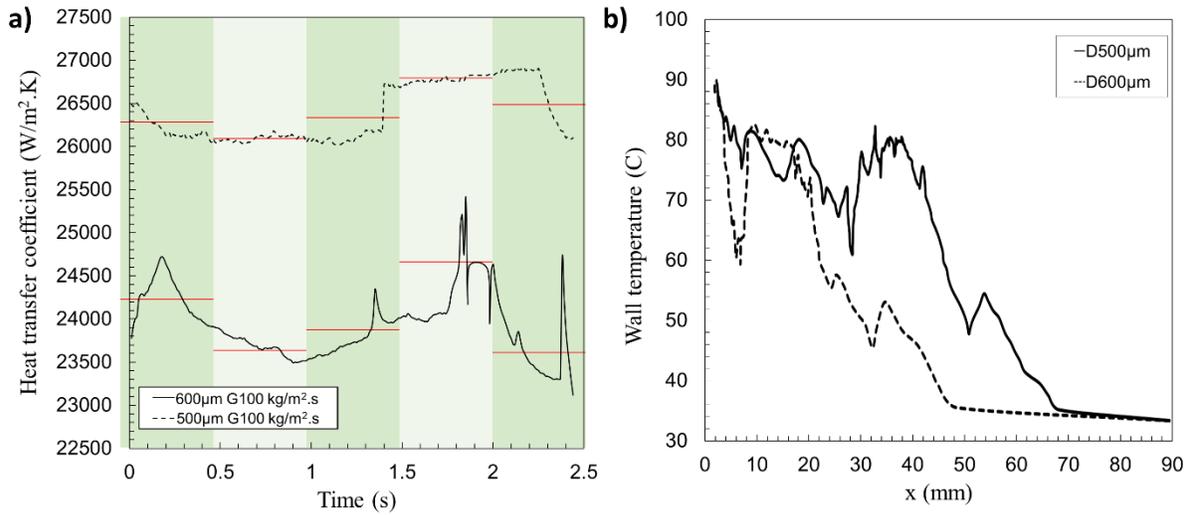

Figure 12. a) Heat transfer coefficient variation with time for 2.5 second period; b) Local wall temperatures for microtubes with D=500μm and D=600μm for cooling heat flux of 30 kW/m$^2$

## 5. Conclusion

This study investigates the hydrothermal properties of condensing flow in microtubes with inner diameters of 250, 500, and 600, and 900μm using numerical and experimental methods. Inlet qualities ranging from 0.2 to 0.8, and inlet vapor mass fluxes ranging from 25 to 225 kg/m$^2$.s were used to characterize the interfacial properties of two phase flow to analyze the heat transfer coefficients and pressure drops in the microtubes. Several conclusions are summarized as follows:

(1) The interfacial characteristics of condensing flow in microtubes with hydraulic diameter smaller than 500μm are majorly different from those with D>500μm.

(2) The effect of vapor quality on liquid film thickness is more pronounced at smaller diameters, which indicates the dominancy of interfacial surface tension force at smaller scales. As vapor quality decreases along the microchannel, the liquid film thickness increases and results in wall shear stress reduction.

(3) A linear relationship exists between flow transition position and inlet vapor mass flux. For the same inlet mass flux, microtube diameter decrement results in the extension of transition location towards the outlet.

(4) After break-up from the vapor column, increasing the inlet vapor mass flux or decreasing the wall heat flux results in an enhancement in the volume of separated bubble. The effect of wall heat flux on bubble size increases with the inlet mass flux.

(5) The heat transfer coefficient trends differ for microtubes with $D_h \leq 500\mu m$ and those with $D_h > 500\mu m$. For microtubes with $D_h \leq 500\mu m$, the effect of inlet mass flux on heat transfer coefficient decreases with steam flow rate, indicating that the convection effect on condensation heat transfer decreases with steam mass flux.

(6) The heat transfer coefficient in the microtube with an inner diameter smaller than 600 µm is more stable compared to the microtube with a larger diameter. As the hydraulic diameter decreases, the annular flow regime extends inside the microtube, which results in a more stable liquid-vapor interface.

**Acknowledgements**

This study was supported by TUBITAK (The Scientific and Technological Research Council of Turkey) Support Program for Scientific and Technological Research Project Grant No. 120M659. The equipment and characterization support are provided by the Sabanci University Nanotechnology Research and Applications Center (SUNUM). The author also thanks Shaghayegh Saeidiharzand for her assistance in capturing Scanning Electron Microscopy (SEM) images.